\documentclass[10pt,aps,prd,nofootinbib,reprint]{revtex4-1}

\usepackage[utf8]{inputenc}
\usepackage{amsmath}
\usepackage{graphicx}
\usepackage{xcolor}
\usepackage{pifont}
\usepackage{cancel}
\usepackage{bbm}
\usepackage{array}
\newcolumntype{H}{>{\setbox0=\hbox\bgroup}c<{\egroup}@{}}
\usepackage{numprint}

\usepackage[colorlinks,pdfusetitle]{hyperref}
\hypersetup{colorlinks=true,allcolors=[rgb]{1,0.56,0}}

\newcommand{\orcid}[1]{\href{https://orcid.org/#1}{#1}}
\newcommand{\e}[1]{\times10^{#1}}

\newcommand{\ts}{\Delta TS}

\begin{document}

\title{Tau Neutrino Identification in Atmospheric Neutrino Oscillations\\Without Particle Identification or Unitarity}

\author{Peter B.~Denton}
\email{pdenton@bnl.gov}
\thanks{\orcid{0000-0002-5209-872X}}
\affiliation{High Energy Theory Group, Physics Department, Brookhaven National Laboratory, Upton, NY 11973, USA}

\date{December 15, 2021}

\begin{abstract}
The largest tau neutrino dataset to date is IceCube's atmospheric tau neutrino appearance dataset containing $>1,000$ tau neutrino and antineutrino events as determined by a fit to a standard three-flavor oscillation framework.
On an event-by-event basis, however, it is impossible to know that any given event is a tau neutrino as they are identical to either an electron neutrino charged-current event or a neutral-current interaction of any active flavor.
Nonetheless, we conclusively show that, using only the cascade sample even without knowledge of the oscillation parameters and without assuming that the lepton mixing matrix is unitary, tau neutrino identification is still possible and there is no viable scenario in which all of the tau neutrino candidates are actually electron neutrinos.
This is primarily due to the matter effect and the tau lepton production threshold, as well as the fact that tau neutrinos are systematically reconstructed at a lower energy than electron neutrinos due to one or more outgoing neutrinos.
This conclusively shows that it is possible for an atmospheric neutrino oscillation experiment to confirm that $U_{\tau1}$, $U_{\tau2}$, and $U_{\tau3}$ are not all zero even with limited particle identification.
\end{abstract}

\maketitle

\section{Introduction}
The tau neutrino is the least-studied particle in the Standard Model (SM) making it a crucial target for additional study.
The global dataset of tau neutrinos is small with the initial data coming from the discovery experiment DONuT \cite{Kodama:2000mp} which detected tau neutrinos via direct production and OPERA \cite{Agafonova:2018auq} which measured tau neutrino appearance in a muon neutrino beam.
In 2010 it was pointed out that IceCube would detect tau neutrinos in their atmospheric dataset \cite{Giordano:2010pr}.
Since then, Super-KamiokaNDE (SK) and IceCube \cite{Li:2017dbe,Aartsen:2019tjl} have both reported detections of tau neutrino appearance in atmospheric neutrinos.
In addition, IceCube has also reported the detection of tau neutrinos from astrophysical sources \cite{Abbasi:2020zmr}.
\begin{center}
\begin{tabular}{c|c|c}
Experiment&Source&$\sim$Events detected\\\hline
DONuT&Production&7.5\\
OPERA&Long-baseline&8\\
SK&Atmospheric&291\\
IceCube&Atmospheric&1804\\
IceCube&Astrophysical&2
\end{tabular}
\end{center}

DONuT, OPERA, and IceCube's astrophysical analyses leveraged the short, but detectable, lifetime of the tau lepton.
SK's atmospheric analysis uses a neural network which leverages tau lepton decay information as well as the various oscillation-related effects discussed in this paper.
Unlike SK, however, the IceCube detector is not sensitive enough to suss out the distinction between tau decays and electron neutrino instigated showers, although IceCube could have some sensitivity after further detector upgrades; see \cite{Li:2016kra}.
That is, without assuming knowledge of the oscillation parameters, it is very difficult or impossible to confirm that the probability of a single event to be a tau neutrino is nonzero.

As the atmospheric oscillation parameters are moderately well understood with independent confirmations of the parameters from long-baseline accelerator experiments \cite{Adamson:2020ypy,T2K:2021xwb,Acero:2019ksn}, robustly testing the oscillation picture is key to confirming that we understand the neutrino sector.
The unitarity framework is a relatively model-independent framework to quantify deviations from the standard three-flavor oscillation picture.
Investigations into this framework have found that the electron neutrino row is fairly well constrained, the muon neutrino row is relatively well constrained, and the tau neutrino row is largely unconstrained using a subset of the above datasets \cite{Parke:2015goa,Ellis:2020hus,Hu:2020oba}.
A new study focused on the tau neutrino row has showed that more datasets than previously used significantly improves the constraints on the tau neutrino row \cite{Denton:2021mso}.

While the atmospheric datasets have vastly more statistics than the others, they have worse event-by-event detection capabilities than the other channels, in particular for IceCube, making it appear as though positive identification of tau neutrinos is impossible.
Nonetheless, we show here that it is possible to confirm the detection of tau neutrinos in the cascade event sample \emph{without knowing the oscillation parameters} and \emph{without assuming the lepton mixing matrix is unitary}.

In order to understand how tau neutrino appearance can be positively determined without assuming unitarity\footnote{Tau neutrino appearance can be easily confirmed if unitarity is confirmed, see e.g.~\cite{Stanev:1999ki,Martinez-Soler:2021sir}.}, we investigate the relevant effects.
Each tau neutrino charged-current (CC) event at IceCube is indistinguishable from an electron neutrino CC event\footnote{Except those when the tau lepton decays to a muon.}.
Thus it might appear that, for the correct oscillation parameters, it is possible that every tau neutrino event could be identified as an electron neutrino event.
This naturally leads one to investigate if IceCube can actually identify the presence of tau neutrinos in their detector without assuming knowledge of the oscillation parameters.
That is, while it is known from long-baseline accelerator experiments that since $\theta_{23}\sim45^\circ$ \cite{Adamson:2020ypy,T2K:2021xwb,Acero:2019ksn}, if one does not assume the $3\times3$ lepton mixing matrix is unitary, then on an event-by-event basis every tau neutrino would appear to be indistinguishable from an electron neutrino.
These various datasets and their impact on the tau row of the lepton mixing matrix without the assumption of unitarity are discussed in \cite{Denton:2021mso} which significantly expands the tau neutrino input beyond previous unitarity analyses \cite{Parke:2015goa,Ellis:2020hus,Hu:2020oba}.

In this paper we prove that it is possible to confirm the existence of tau neutrinos in atmospheric neutrino oscillations without assuming the lepton mixing matrix is unitary and without identifying the specifics of the tau lepton's hadronic decays or by measuring its lifetime.
We make the absolute minimal number of assumptions possible and quantify the impact of each assumption on the capability to identify the existence of tau neutrinos in the atmospheric neutrino flux.
We assume that neutrinos oscillate, neutrinos experience the matter effect \cite{Wolfenstein:1977ue}, the large tau lepton mass gives rise to a threshold effect, and tau neutrinos deposit systematically less energy than an electron neutrino with the same energy due to decays to undetected neutrinos.
The combination of these effects provides enough information to confirm that tau neutrinos are detected in IceCube's atmospheric data without any prior knowledge on the oscillation parameters.
While the analysis presented here is focused on atmospheric neutrinos at IceCube, the story is equivalent for SK as well as future atmospheric neutrino experiments such as Hyper-KamiokaNDE \cite{Abe:2018uyc}, KM3NeT/ORCA \cite{Adrian-Martinez:2016fdl}, DUNE \cite{Abi:2020evt}, ICAL at INO \cite{Kumar:2017sdq}, and others.

This paper is organized as follows.
First, in section \ref{sec:uv} we discuss how we will apply in a self-consistent fashion a unitarity violation framework including the matter effect.
This gives us context to test whether or not tau neutrinos are detected at all while still working within an oscillation framework.
Second, we describe how we simulate an atmospheric neutrino experiment in section \ref{sec:atm}.
Third, we present our results in section \ref{sec:results} along with some interpretations in section \ref{sec:discussion}.
Finally, we conclude and summarize the paper in section \ref{sec:conclusions}.

\section{Unitarity Violation Overview}
\label{sec:uv}
In order to quantify the ability to differentiate scenarios with and without tau neutrinos, we consider unitary violation (UV) for the case without tau neutrinos.
UV of the $3\times3$ matrix that is probed in most neutrino oscillation experiments is a generic framework to parameterize various BSM scenarios, often related to neutrino mass generation.
Most simplistically, any additional sterile neutrino state, whether connected to neutrino mass generation \cite{Minkowski:1977sc,Schechter:1980gr,Foot:1988aq} or otherwise is expected to lead to apparent UV.
Additionally, scenarios involving neutrinos propagating in extra dimensions \cite{ArkaniHamed:1998vp,ArkaniHamed:1998sj,Bhattacharya:2009nu} would also appear as UV.
We focus on scenarios that are parameterized as additional gauge singlet fermions that may or may not be kinematically accessible in a given experimental configuration.

Specifically, we parameterize a UV scenario with two numbers:
\begin{itemize}
\item $n$: the total number of neutrinos,
\item $m$: the number of neutrinos that are kinematically accessible.
\end{itemize}
While many possible combinations are viable, we focus on the three most interesting pairs $(n,m)$: (3,3), (4,4), and (5,3).
The (3,3) scenario is the usual standard three-flavor oscillation scenario.
The (4,4) scenario is the case with one sterile neutrino that is light enough to be kinematically produced ultrarelativisitically ($m_4\lesssim10$ keV for neutrinos from neutrons and $\lesssim15$ MeV for neutrinos from pions and muons) but heavy enough that the oscillations cannot be directly probed ($m_4\gtrsim10$ eV).
The (5,3) scenario is the case with two sterile neutrinos that are heavy enough to be not produced ($m_4\gtrsim40$ MeV) and can be directly related to the minimal unitary violation scheme often parameterized via the $\alpha$ matrix \cite{Blennow:2016jkn}.
For the case with heavy neutrinos we focus on the (5,3) case instead of (4,3) or (6,3) because with two additional neutrinos there are enough degrees of freedom to completely cover all available degrees of freedom in the $3\times3$ matrix.
That is, in principle, given enough high-precision measurements of each different oscillation, channel, one could differentiate the (4,3) scenario from the (5,3) scenario, but not the (5,3) scenario from the (6,3) or larger scenario.
Additional scenarios such as (5,4) are different from (5,3) due to the matter effect, but we focus on these three cases for concreteness.

In any of the above-mentioned scenarios, if any of the new neutrinos masses are close to the kinematic limit or are accessible in some experiments and not others, or if the new masses are close enough to each other to induce their own oscillations, then the situation becomes considerably more involved.
We avoid these regions since the UV framework is designed to be relatively independent of the details of the sterile neutrinos and the regions of parameter space discussed here cover much of the phenomenologically and theoretically interesting parameter space.

\subsection{Probability}
If $n=m$ then the oscillation probabilities can be calculated in the usual fashion \cite{Zaglauer:1988gz,Ohlsson:1999um,Kimura:2002wd,Li:2018ezt,Denton:2019ovn} since the full unitary matrix $U$ is only composed of accessible states.
If $n>m$ then some care is needed, see \cite{FernandezMartinez:2007ms,Fong:2017gke}.
We define $N$ as the $m\times m$ submatrix composed of the first $m$ columns and rows\footnote{While the first $m$ columns clearly must be included, the choice of rows appears to be somewhat more arbitrary, so long as it includes the three active flavors.
In fact, in the (5,4) case for example, instead of $N$ being a $4\times4$ matrix one could have $N$ as a $3\times4$ matrix and the matter potential matrix as only $3\times3$ composed of the active states.
Alternatively, either one of the two sterile flavors could be added to the bottom of $N$ which would change nothing since the matter potential would gain a zero which would annihilate the sterile state anyway.} of the larger unitary matrix, $U$.
The Hamiltonian in the mass basis that describes propagation is the usual one extended for $m$ accessible neutrinos,
\begin{multline}
H_{\rm mass}=\frac1{2E}
\begin{pmatrix}
0\\
&\Delta m^2_{21}\\
&&\Delta m^2_{31}\\
&&&\Delta m^2_{41}\\
&&&&\ddots
\end{pmatrix}\\
+N^T
\begin{pmatrix}
V_{\rm CC}+V_{\rm NC}\\
&V_{\rm NC}\\
&&V_{\rm NC}\\
&&&0\\
&&&&\ddots
\end{pmatrix}N^*\,,
\end{multline}
where $V_{\rm CC}=\sqrt2G_FN_e$, $V_{\rm NC}=-\frac1{\sqrt2}G_FN_n$, and $N_f$ is the number density of matter fermion $f$.
Note that while $m_1^2\mathbbm1$ can be subtracted from the Hamiltonian without changing any oscillation effects since we are in the mass basis, $V_{\rm NC}\mathbbm1$ cannot be since $N$ is not unitary.

In general, we find that the oscillation amplitude is \cite{Giunti:2004zf,Antusch:2006vwa,FernandezMartinez:2007ms},
\begin{equation}
\mathcal A_{\alpha\beta}=\frac{[N^*We^{-i\Lambda L}W^\dagger N^T]_{\alpha\beta}}{\sqrt{(NN^\dagger)_{\alpha\alpha}(NN^\dagger)_{\beta\beta}}}\,,
\label{eq:A UV}
\end{equation}
where $W$ is the $m\times m$ unitary matrix composed of eigenvectors of the Hamiltonian and $\Lambda$ is the diagonal matrix composed of the eigenvalues of the Hamiltonian.
Quantities like $(NN^\dagger){\alpha\alpha}$ are $\sum_{i=1}^m|N_{\alpha i}|^2$.
The probability is then $P_{\alpha\beta}=|\mathcal A_{\alpha\beta}|^2$.
In vacuum $N_e=N_n=0$ and thus $\lambda_i=m_i^2/2E$ and $W=\mathbbm1$.
In addition, if $m=n$ then the denominator in eq.~\ref{eq:A UV} is 1 since $N$ is unitary in this case and we recover the usual expression for the oscillation probabilities.

The denominator accounts for production and detection effects while the terms in the numerator include the propagation in matter.
We assume that any flavor or mass state is either produced with no kinematic suppression or is completely disallowed, with the exception of the tau neutrino for which the threshold effects, parameterized with $R_{\tau\mu}$, are discussed in section \ref{sec:atm} below.

For concreteness we parameterize the mixing matrix as $U_{35}U_{25}U_{15}U_{34}U_{24}U_{14}U_{23}U_{13}U_{12}$ where we have not included $U_{45}$ as we already have enough degrees of freedom and this rotation will not affect oscillations, for more on the choice parameterization, see \cite{Denton:2020igp}.
In order to turn off tau neutrino appearance we set the mixing angles as described in table \ref{tab:no tau angles}.

\begin{table}
\centering
\caption{The angles for each of the mixing matrix configurations such that there is no tau neutrino appearance.
That is, for each of the listed matrix setting the angles to the listed value sets $U_{\tau1}=U_{\tau2}=U_{\tau3}=0$.}
\label{tab:no tau angles}
\begin{tabular}{l|c|c|c}
&(3,3)&(4,4)&(5,3)\\\hline
Angles set to $90^\circ$&$\theta_{23}$&$\theta_{34}$&$\theta_{35}$\\
Angles set to $0^\circ$&$\theta_{13}$, $\theta_{i4}$, $\theta_{i5}$&$\theta_{i4}$, $\theta_{i5}$&$\theta_{i5}$
\end{tabular}
\end{table}

\subsection{Flux and Cross Section}
In addition to modifying the probability, the flux and cross section are changed relative to the SM expectation in these various UV schemes.
The true cross section and flux for each neutrino flavor are related to the SM cross section and flux by \cite{Antusch:2006vwa}
\begin{align}
\sigma_{{\rm CC},\alpha}&=\sigma_{{\rm CC},\alpha}^{\rm SM}(NN^\dagger)_{\alpha\alpha}\,,\\
\Phi_\alpha&=\Phi_\alpha^{\rm SM}(NN^\dagger)_{\alpha\alpha}\,.
\end{align}
So the true flux and cross section are either the same as the SM ($m=n$) or less ($m<n$).

Note that these corrections exactly cancel those in the denominator of eq.~\ref{eq:A UV}.
So most experiments (including atmospheric neutrino experiments) are mostly sensitive to a reduced probability,
\begin{equation}
P_{\alpha\beta}^r=\left|[N^*We^{-i\Lambda L}W^\dagger N^T]_{\alpha\beta}\right|^2\,.
\label{eq:Pr UV}
\end{equation}

In addition, $G_F$ in the matter effect must also be corrected from the measured value $G_F^M$ since it is measured from muon decays to two neutrinos via
\begin{equation}
G_F=\frac{G_F^M}{\sqrt{(NN^\dagger)_{ee}(NN^\dagger)_{\mu\mu}}}\,.
\end{equation}
We can see that in the (5,3) case $G_F\ge G_F^M$ while in the (3,3) and (4,4) cases $G_F=G_F^M$.

\begin{figure*}
\centering
\includegraphics[width=0.49\textwidth]{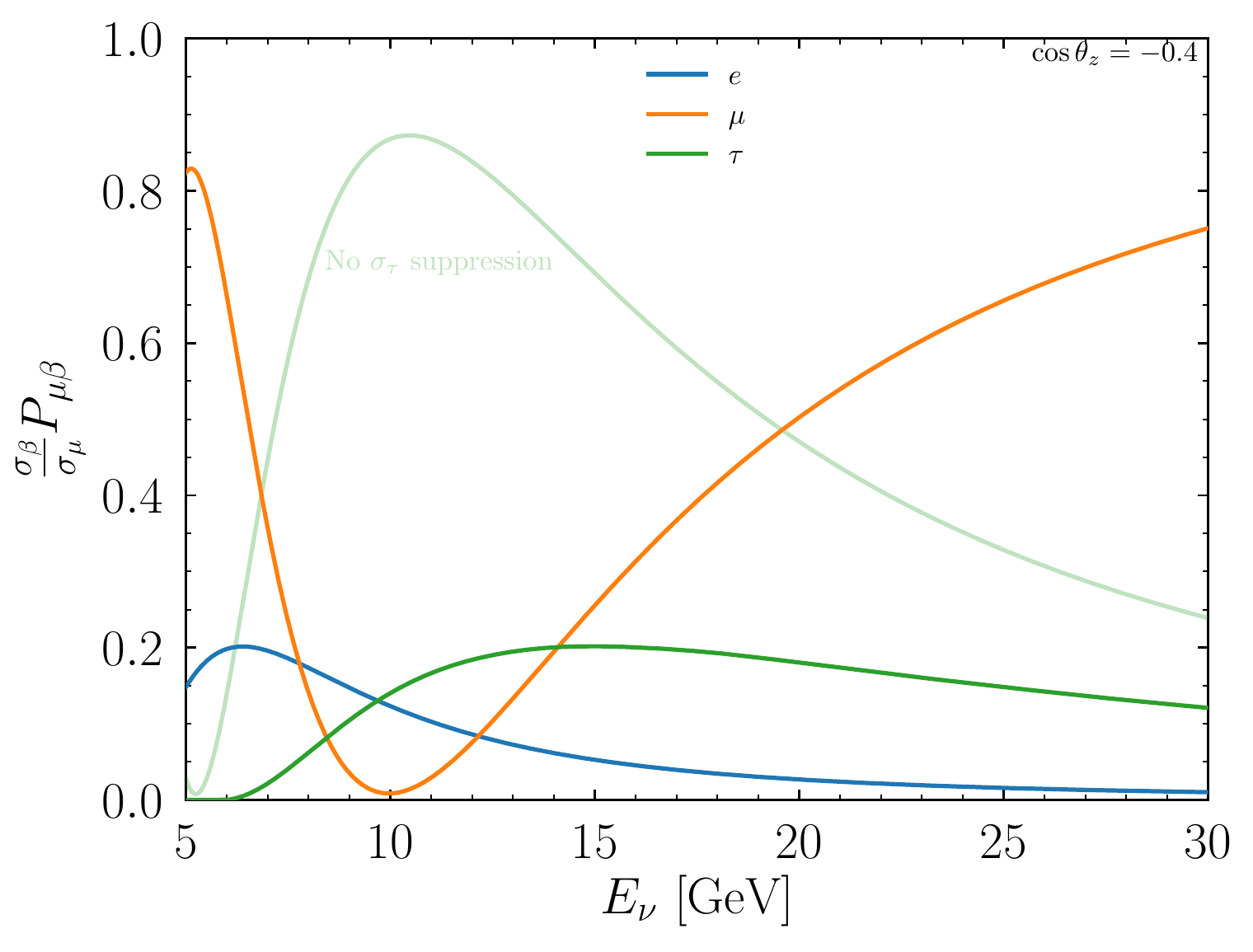}
\includegraphics[width=0.49\textwidth]{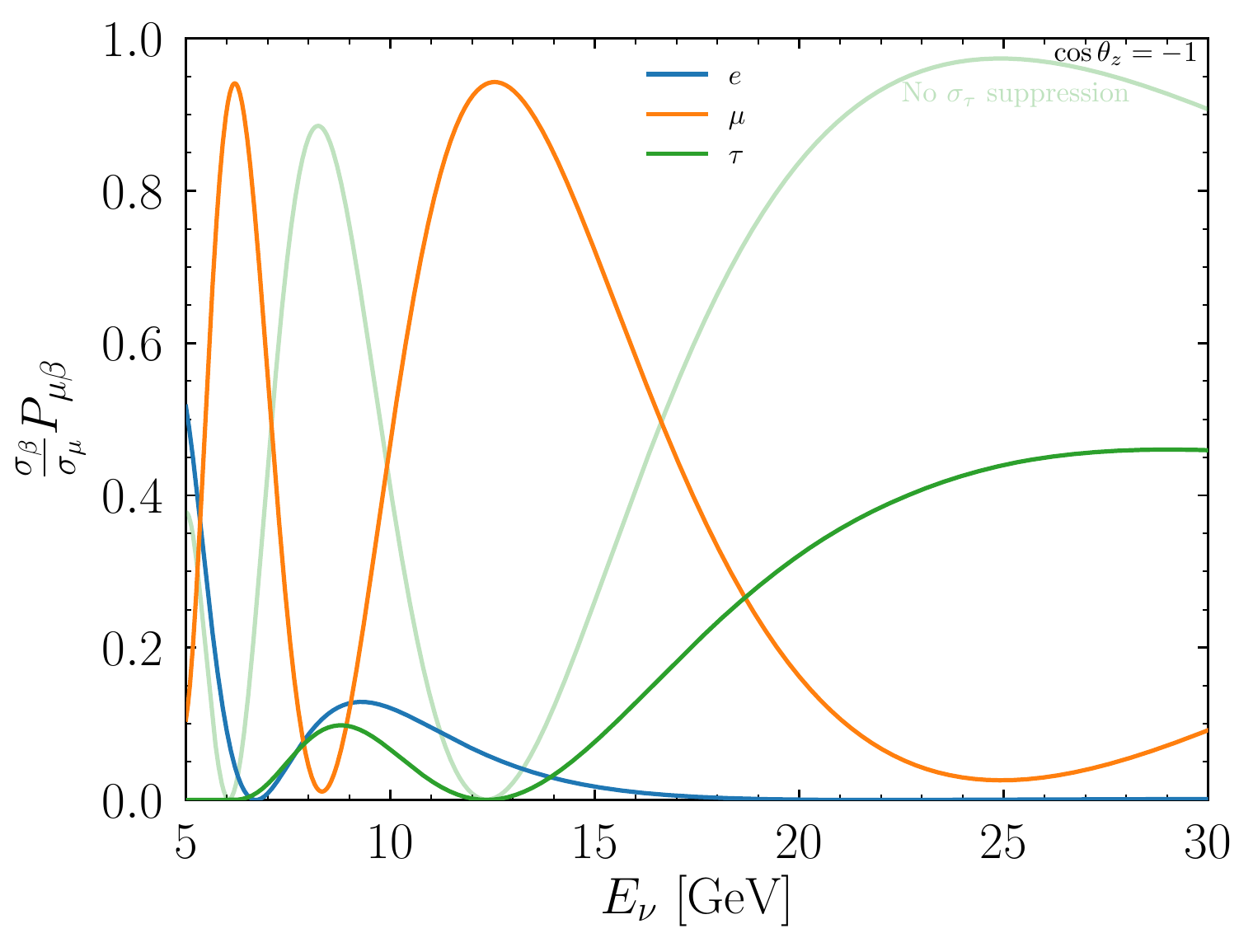}
\caption{The atmospheric probability with the tau production threshold as a function of the true neutrino energy.
On the left is for events through the mantle and the right is core-crossing events.
In light green is tau neutrino appearance probability without the tau production threshold effect.}
\label{fig:atm prob}
\end{figure*}

\section{Atmospheric Tau Neutrino Appearance Detection}
\label{sec:atm}
The atmospheric neutrino flux is dominated by muon neutrinos during production with subleading contributions from muon antineutrinos \cite{Honda:2015fha}.
Under the standard oscillation picture, using the well-established result that $\theta_{23}\sim45^\circ$ and $\Delta m^2_{32}\sim2.5\e{-3}$ eV$^2$, a significant number of muon neutrinos disappear at various energies and zenith angles and baselines across the range 5-50 GeV, and most of these have oscillated into tau neutrinos; see fig.~\ref{fig:atm prob} which shows the $P(\nu_\mu\to\nu_\alpha)$ probabilities rescaled by their cross sections.
Note that while in the left panel at $\cos\theta_z=-0.4$ the $\nu_\mu\to\nu_\tau$ curve could be partially replicated by the $\nu_\mu\to\nu_e$ curve with a higher value of $\Delta m^2_{32}$, this same shift would not work for core-crossing $\cos\theta_z=-1$ events shown in the right panel.
In order to confirm that tau neutrinos can be (and thus have been given the existing IceCube and SK's current datasets) detected, we compare the signal from the expected flux to one with no tau neutrinos.
We parameterize this lack of tau neutrinos in a unitarity-violating scheme described in section \ref{sec:uv}.
That is, the detection of tau neutrinos provides an important probe on the tau row unitarity, the least well-constrained part of the lepton mixing matrix \cite{Parke:2015goa,Ellis:2020hus,Hu:2020oba}.

In this section we demonstrate how an atmospheric neutrino experiment, such as IceCube, can probe the tau neutrino row of the lepton mixing matrix, without the assumption of unitarity, and without directly identifying the outgoing tau lepton.
IceCube classifies events into tracks and cascades \cite{Aartsen:2019tjl}.
Tracks are longer cylindrical events that come from muon neutrino CC events as well as tau neutrino CC events where the tau decays to a muon.
These events have very good angular resolution but often poor energy resolution.
Cascades are roughly spherical events that include all the remaining channels: electron neutrino CC events, tau neutrino CC events so long as the tau lepton decays hadronically or to an electron, and NC events.
Cascade events have excellent energy resolution but often poor angular resolution.
It would appear that, without the constraint of unitarity relating tau neutrino appearance to the other oscillation channels, that since every tau neutrino cascade event could be reclassified as an electron neutrino event, the data would be indistinguishable from the case where there was no tau neutrino appearance.
In this section, we show that this is not true due to properties of the tau lepton and the matter effect.

In order to illustrate this, we suppose that an experiment performs a measurement of atmospheric neutrinos and measures something consistent with known oscillation parameters.
Then we show that it is impossible to reproduce this data under the assumption of no tau neutrino appearance in the various UV configurations described in section \ref{sec:uv}.
Finally, in an effort to understand which component contributes to the tau neutrino identification, we turn on and off various known effects such as the matter effect or the tau production threshold.

Our setup is a version of an IceCube analysis \cite{Aartsen:2019tjl} containing all of the relevant features for tau neutrino identification.
We conservatively focus on the cascade component as the track events contain a relatively smaller amount of information for tau neutrino appearance.
We also only include up-going events as down-going events similarly have less information and more complicated backgrounds.
We use the same binning as IceCube: 5 uniform bins in $\cos\theta_z<0$ and 8 energy bins from $5.6$ GeV to $56$ GeV distributed logarithmically.
We also assume that the initial atmospheric flux is dominantly $\nu_\mu$ and is a single power law $\Phi_i(E_{\rm true})\propto E_{\rm true}^{-\gamma}$ for $\gamma=3$, generally consistent with the predicted atmospheric flux over the energy range in question \cite{Honda:2015fha}.
We ignore track -- cascade misidentification.
Finally, we work in the parameter space of the $E_{\rm reco}$ and $\cos\theta_z$ parameters and assume that fluctuations in the various parameters are small compared to the bin sizes.
This allows us to write down how oscillations and other cross-section-related features modify the detected cascade flux spectrum $d^2N/dE_{\rm reco}d\cos\theta_z$ relative to the initial atmospheric flux times the total neutrino-nucleon cross section, $\Phi_i(E_{\rm reco})\sigma_{\rm tot}(E_{\rm reco})$,
\begin{widetext}
\begin{multline}
\mathcal R_c(E_{\rm reco},\cos\theta_z)\equiv
\frac{\frac{d^2N_c}{dE_{\rm reco}d\cos\theta_z}}{\Phi_i(E_{\rm reco})\sigma_{\rm tot}(E_{\rm reco})}=
f_{\rm CC}\left[P_{\mu e}^r(E_{\rm reco},\cos\theta_z)\right.\\
\left.+(1-f_{\tau\mu})\eta_{\nu_\tau}^{\gamma-1}R_{\tau\mu}(E_{\rm reco}/\eta_{\nu_\tau})P_{\mu\tau}^r(E_{\rm reco}/\eta_{\nu_\tau},\cos\theta_z)\right]
+(1-f_{\rm CC})\eta_{\rm NC}^{\gamma-1}\sum_{\beta\in\{e,\mu,\tau\}}P_{\mu\beta}^r(E_{\rm reco}/\eta_{\rm NC},\cos\theta_z)\,,
\label{eq:IC cascade spectrum}
\end{multline}
\end{widetext}
where $f_{\rm CC}\simeq0.7$ is the CC cross section fraction, $f_{\tau\mu}\simeq0.174$ is the fraction of tau leptons that decay to muons, $\eta_{\nu_\tau}\simeq0.625$ is the ratio of the tau neutrino's reconstructed energy divided by its true energy \cite{Aartsen:2019tjl}, $R_{\tau\mu}$ is the ratio of $\nu_\tau-N$ cross section to $\nu_\mu$ cross section from \cite{Jeong:2010nt}, and $\eta_{\rm NC}\simeq\frac13$ is the fraction of energy deposited in NC interactions.
The two $\eta^{\gamma-1}$ terms include $\gamma$ factors of the energy shifts since the spectrum decreases as the energy increases, and $-1$ factor of the energy shifts since the cross section increases linearly with energy in this energy range \cite{Gandhi:1998ri}.
We define the quantity in eq.~\ref{eq:IC cascade spectrum} as $\mathcal R_c(E_{\rm reco},\cos\theta_z)$.
A similar quantity could also be defined for tracks as well.

\begin{figure*}
\centering
\includegraphics[width=0.49\textwidth]{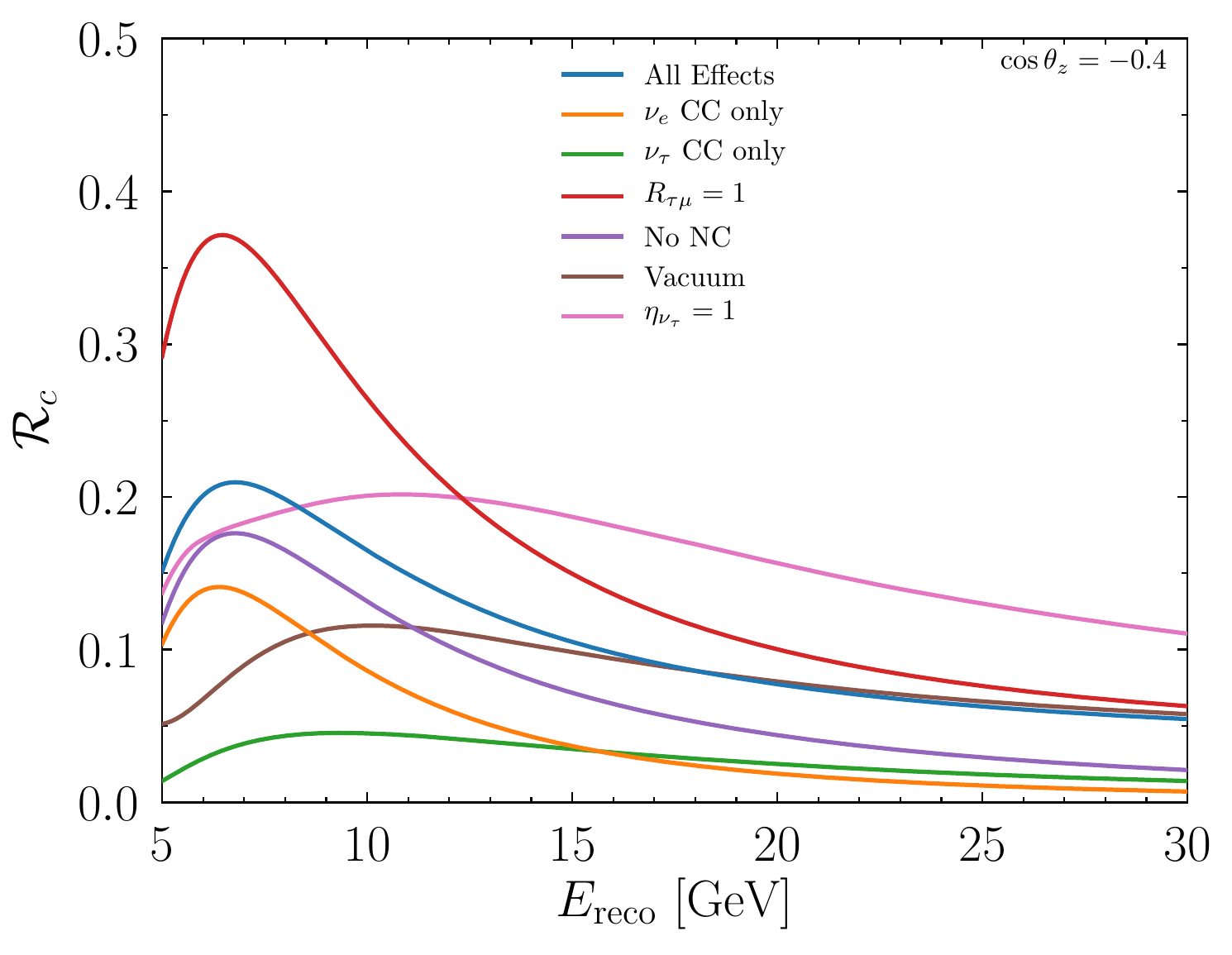}
\includegraphics[width=0.49\textwidth]{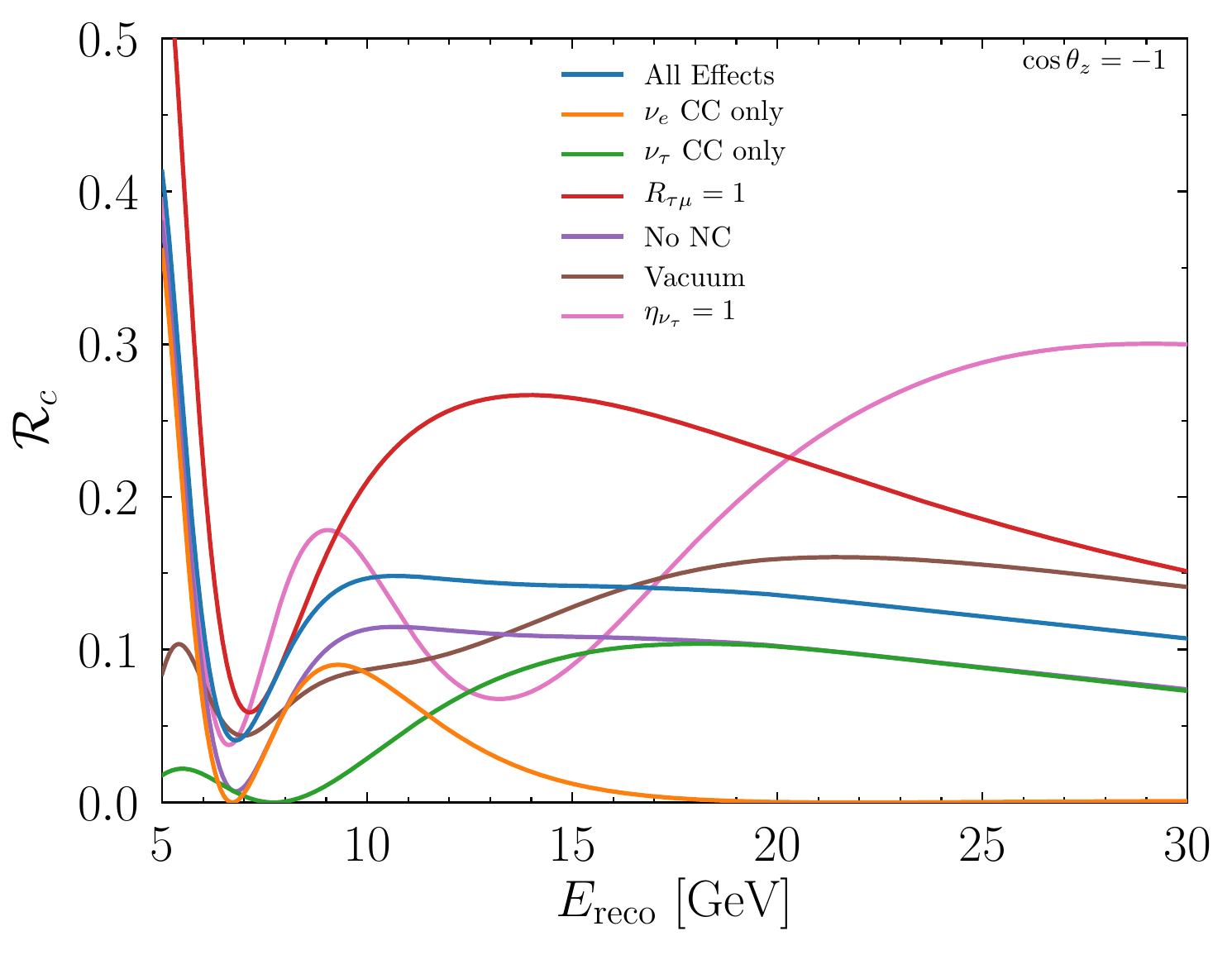}
\caption{The ratio of the cascade flux to the initial flux as a function of the reconstructed energy for different zenith angles.
The different curves represent the inclusion of different effects.
The blue curve is the standard scenario.
The orange curve is the scenario with no tau neutrinos or NC interactions.
The green curve is the scenario with no electron neutrinos or NC interactions.
The red curve is the scenario with no tau lepton threshold effect.
The purple curve is the scenario with no NC interactions.
The brown curve is the scenario with no matter effect.
The pink curve is the scenario where tau neutrinos would be reconstructed with full energy.}
\label{fig:cascade ratio}
\end{figure*}

In fig.~\ref{fig:cascade ratio} we show eq.~\ref{eq:IC cascade spectrum} as a function of reconstructed energy for various different zenith angles and with various different components included.
We see for example at $\cos\theta_z=-0.4$ through the mantle that the scenario with only electron neutrino CC interactions (orange) is quite similar to the standard case (blue) up to a normalization which could be accommodated by an uncertainty in the flux or the mixing parameters.
For more up-going events, however, the shape is no longer correct as there would not be enough events at high energy.
This shows explicitly why tau neutrinos can be identified.

To estimate the information an IceCube-like experiment will measure in each bin, we calculate a series of weights which contain information about cross section, flux, and detector efficiencies.
These weights are defined based on the $S/\sqrt B$ in each bin as provided by IceCube \cite{Aartsen:2019tjl} where $S$ is the signal and $B$ is the background for three years of running.
In bin $j,k$, the weight is,
\begin{multline}
w_{jk}=\left\{\left(\frac S{\sqrt B}\right)_{jk}\right.\\
\left.\times\left[\int d(\cos\theta_z)\int dE_{\rm reco}\mathcal R_c(E_{\rm reco},\cos\theta_z)\right]^{-1}\right\}^2\,,
\label{eq:weight}
\end{multline}
where the integrals are over the area of the bin and $\mathcal R_c$ is calculated for the benchmark oscillation parameters and all the other known particle physics where we note that the IceCube data is well described the SM for standard oscillation parameters.

Then we define the test statistic between two hypotheses as
\begin{widetext}
\begin{equation}
\ts=\sum_{j,k}w_{jk}\left\{\int d(\cos\theta_z)\int dE_{\rm reco}\left[\mathcal R_c(E_{\rm reco},\cos\theta_z)_{\nu_\tau}-\mathcal R_c(E_{\rm reco},\cos\theta_z)_{\cancel{\nu_\tau}}\right]\right\}^2\,,
\label{eq:chisq}
\end{equation}
\end{widetext}
where the sum is over the $\cos\theta_z$ and $E_{\rm reco}$ bins and in the $\nu_\tau$ term the oscillation parameters are fixed to the standard oscillation parameters while in the $\cancel{\nu_\tau}$ term the oscillation parameters are allowed to float, but there are no tau neutrinos as described in table \ref{tab:no tau angles} for the different matrix scenarios.
We also allow for a free flux normalization in the $\cancel{\nu_\tau}$ term to conservatively account for the uncertainty in the atmospheric flux.
While eq.~\ref{eq:chisq} behaves as a $\Delta\chi^2$, the numerical value should not be interpreted too strongly as a sensitivity as we have made several simplifications.
Nonetheless, it is sufficient to show the relative impact of various components of eq.~\ref{eq:IC cascade spectrum} on the ability to disfavor no tau neutrino appearance due to $U_{\tau1}=U_{\tau2}=U_{\tau3}=0$.
While in eq.~\ref{eq:weight} we always include all relevant physics parameters, in eq.~\ref{eq:chisq} we will sometimes turn some of them off for both terms.
We will use eq.~\ref{eq:chisq} to compare the cases of the standard oscillation parameters and that where the first three elements in the tau neutrino row are zero.

\section{Results}
\label{sec:results}
In table \ref{tab:chisq} we show all of the numerical results for every combination of physical processes as well as every combination of electron neutrino row constraints and $\Delta m^2_{31}$ constraints.
In the first four columns green check marks refer to the SM version of the mentioned effect.
The nature of the numerical results are summarized in this section.

\definecolor{forestgreen}{rgb}{0.13, 0.55, 0.13}
\definecolor{lgreen}{rgb}{0.7, 0.9, 0.7}
\def\y{\textcolor{forestgreen}{\text{\ding{51}}}}
\def\n{\textcolor{red}{\text{\ding{55}}}}
\begin{table*}
\centering
\caption{Test statistic disfavoring $U_{\tau1}=U_{\tau2}=U_{\tau3}=0$ relative to the standard case in three different matrix configurations without the electron neutrino row constrained (left) and with the electron neutrino row constraint (right).
The electron neutrino row constraint makes the (3,3) case overdetermined and we do not consider it on the left.
Note that the best fit values of the oscillation parameters are often extremely far away from known values from other neutrino oscillation measurements.
See the text for remaining details.}
\npdecimalsign{.}
\nprounddigits{0}
\begin{tabular}{c|c|c|c|c||n{4}{0}|n{4}{0}|n{4}{0}}
\multicolumn{8}{c}{No $\nu_e$ row constraint}\\
NC&Matter&$\eta_{\nu_\tau}$&$R_{\tau\mu}$&$\Delta m^2_{31}$ const.&{(3,3)}&{(4,4)}&{(5,3)}\\\hline\hline
\n&\n&\n&\n&\n&1580.1000&0.0017&412.7990\\
\n&\n&\n&\n&\y&1580.1000&0.0020&471.3640\\
\n&\n&\n&\y&\n&358.4090&15.6443&56.7367\\
\n&\n&\n&\y&\y&358.4090&27.1838&65.0166\\\hline
\n&\n&\y&\n&\n&129.1190&0.3870&49.5247\\
\n&\n&\y&\n&\y&129.1190&0.7911&0.5577\\
\n&\n&\y&\y&\n&52.2984&0.4631&1.0416\\
\n&\n&\y&\y&\y&52.2984&0.5805&0.6034\\\hline
\n&\y&\n&\n&\n&1323.5600&497.2170&702.3680\\
\n&\y&\n&\n&\y&1323.5500&1013.6800&702.3680\\
\n&\y&\n&\y&\n&283.2720&117.5280&123.5460\\
\n&\y&\n&\y&\y&283.2710&231.4810&123.5470\\\hline
\n&\y&\y&\n&\n&109.3930&58.1574&95.2684\\
\n&\y&\y&\n&\y&109.3910&89.7096&94.6506\\
\n&\y&\y&\y&\n&34.2101&30.5327&23.2745\\
\n&\y&\y&\y&\y&34.2098&30.5323&29.1739\\\hline
\y&\n&\n&\n&\n&726.8320&0.0018&433.2670\\
\y&\n&\n&\n&\y&726.8320&0.0020&457.7420\\
\y&\n&\n&\y&\n&155.0700&14.2073&61.4279\\
\y&\n&\n&\y&\y&155.0700&26.6811&69.8274\\\hline
\y&\n&\y&\n&\n&99.2813&0.4177&50.1815\\
\y&\n&\y&\n&\y&99.2813&1.0672&58.9013\\
\y&\n&\y&\y&\n&30.9528&0.4101&11.6824\\
\y&\n&\y&\y&\y&30.9528&0.5810&4.2686\\\hline
\y&\y&\n&\n&\n&972.9480&613.6520&702.3680\\
\y&\y&\n&\n&\y&702.3680&663.7490&699.6800\\
\y&\y&\n&\y&\n&123.5460&118.9910&123.5460\\
\y&\y&\n&\y&\y&123.5460&123.5390&123.4310\\\hline
\y&\y&\y&\n&\n&115.2670&81.9912&54.8504\\
\y&\y&\y&\n&\y&115.2670&61.5077&74.0571\\
\y&\y&\y&\y&\n&34.6758&13.6077&22.6753\\
\y&\y&\y&\y&\y&34.5697&15.1326&16.3260
\end{tabular}
\qquad
\begin{tabular}{c|c|c|c|c||n{4}{0}|n{4}{0}}
\multicolumn{7}{c}{$\nu_e$ row constrained}\\
NC&Matter&$\eta_{\nu_\tau}$&$R_{\tau\mu}$&$\Delta m^2_{31}$ const.&{(4,4)}&{(5,3)}\\\hline\hline
\n&\n&\n&\n&\n&0.0018&409.4410\\
\n&\n&\n&\n&\y&0.0492&427.3570\\
\n&\n&\n&\y&\n&16.6613&62.4232\\
\n&\n&\n&\y&\y&28.0237&62.6925\\\hline
\n&\n&\y&\n&\n&0.3898&55.4897\\
\n&\n&\y&\n&\y&0.9825&58.4214\\
\n&\n&\y&\y&\n&0.5257&13.0388\\
\n&\n&\y&\y&\y&0.6571&13.6102\\\hline
\n&\y&\n&\n&\n&511.1380&702.5160\\
\n&\y&\n&\n&\y&1267.0800&695.8580\\
\n&\y&\n&\y&\n&103.6200&123.5500\\
\n&\y&\n&\y&\y&276.2550&123.7710\\\hline
\n&\y&\y&\n&\n&59.2359&111.0610\\
\n&\y&\y&\n&\y&93.3426&109.2720\\
\n&\y&\y&\y&\n&21.2975&16.5829\\
\n&\y&\y&\y&\y&26.2339&26.4915\\\hline
\y&\n&\n&\n&\n&169.0630&434.8720\\
\y&\n&\n&\n&\y&169.9770&442.1200\\
\y&\n&\n&\y&\n&41.5401&66.5104\\
\y&\n&\n&\y&\y&45.6626&66.4517\\\hline
\y&\n&\y&\n&\n&20.4041&55.6463\\
\y&\n&\y&\n&\y&23.7948&59.6376\\
\y&\n&\y&\y&\n&3.6862&12.7624\\
\y&\n&\y&\y&\y&3.7700&13.6100\\\hline
\y&\y&\n&\n&\n&477.7600&702.4530\\
\y&\y&\n&\n&\y&666.8050&702.4100\\
\y&\y&\n&\y&\n&122.9600&123.5480\\
\y&\y&\n&\y&\y&123.5170&123.5270\\\hline
\y&\y&\y&\n&\n&53.5628&110.8010\\
\y&\y&\y&\n&\y&63.3545&112.2150\\
\y&\y&\y&\y&\n&17.8473&32.6562\\
\y&\y&\y&\y&\y&17.9043&33.1861
\end{tabular}
\label{tab:chisq}
\end{table*}

Since the electron neutrino row is quite well measured, we consider the option to fix it to its best fit values or let it float freely.
Similarly, as a number of experiments have measured consistent values of $\Delta m^2_{31}$ we also either fix it or let it float freely (we always fix $\Delta m^2_{21}$ and have confirmed that its effect on the main results are small).
We vary these assumptions as widely as possible to show where the effects come from.
The different effects considered (affecting both terms in eq.~\ref{eq:chisq} but not in eq.~\ref{eq:weight} which defined the experimental setup) are as follows.
\begin{itemize}
\item \textbf{NC}: Whether or not the NC interaction is included.
That is, no NC term means that the final term in eq.~\ref{eq:IC cascade spectrum} is not included.
This has a very small effect regardless of what other effects are included.
\item \textbf{Matter}: Whether or not the matter effect is included.
Without the matter effect $V_{\rm CC}=V_{\rm NC}=0$.
\item $\boldsymbol{\eta_{\nu_\tau}}$: Whether or not the tau neutrino reconstruction effect is taken into account.
With this effect set to the SM this parameter is taken to be 0.625, without it, it is taken to be 1.
\item $\boldsymbol{R_{\tau\mu}}$: Whether or not the tau production threshold effect is taken into account.
With this effect the physical values for $R_{\tau\mu}<1$ are taken, without it, $R_{\tau\mu}=1$ for all neutrino energies.
\item $\boldsymbol{\Delta m^2_{31}}$ \textbf{constraint}: Whether or not $\Delta m^2_{31}$ is fixed to its best fit value.
Note that $\Delta m^2_{21}$ is always fixed to its best fit value; varying it does not have a significant effect on the results.
\item \textbf{Electron neutrino row constraint}: Whether or not $U_{e1}$, $U_{e2}$, and $U_{e3}$ are fixed to their best fit values.
While there is some potential hints of unitary violation in the electron neutrino row \cite{Giunti:2010zu,Mention:2011rk,Aguilar:2001ty,Aguilar-Arevalo:2020nvw}, we use fits not including those datasets as at least some of them may have nonoscillation explanations.
We do not include any hints for unitary violation in the electron row which is equivalent to fixing $\theta_{1j}$ to their best fit values.
\end{itemize}

We could also consider fixing the muon neutrino row of the mixing matrix to the best fit value.
We do not include this scenario as we see sensitivity to identifying tau neutrinos without fixing it.
In addition, it is rather less well constrained than the electron neutrino row \cite{Parke:2015goa,Ellis:2020hus,Hu:2020oba} so letting it float freely, while conservative, is not exceedingly so.

Many of the best fit scenarios have parameters that are extremely inconsistent with other oscillation measurements.
Nonetheless, we see that a measurement of atmospheric neutrino cascades \emph{alone} in this energy and zenith angle parameter space allows for a determination of the parameters in the tau neutrino row.

\section{Discussion}
\label{sec:discussion}
We calculated the test statistic in many cases; we now distill the salient features.
First, without any of the features listed above -- the NC events, the matter effect, the tau neutrino reconstruction effect, and the tau lepton production threshold: the (\n,\n,\n,\n,\n) row on the right of table \ref{tab:chisq} -- we find that it is impossible to identify tau neutrinos even if we assume we know the electron neutrino row and $\Delta m^2_{31}$.
This confirms our expectations that one can dial up the $\nu_\mu\to\nu_e$ oscillation probability and exactly compensate for the missing tau neutrinos.

Second, with none of the effects included and no constraints on the electron neutrino row or $\Delta m^2_{31}$: the (\n,\n,\n,\n,\n) row on the left of table \ref{tab:chisq}, in the (3,3) case there is a large sensitivity.
This is because the only free parameter, other than $\Delta m^2_{31}$, is $\theta_{12}$ so the only way that cascades would appear is via $\Delta m^2_{21}$ oscillations to electron neutrinos which does not replicate the observed cascades due to tau neutrinos.
This is one scenario where allowing $\Delta m^2_{21}$ to float would provide a slightly different result.

\begin{figure*}
\includegraphics[width=0.32\textwidth]{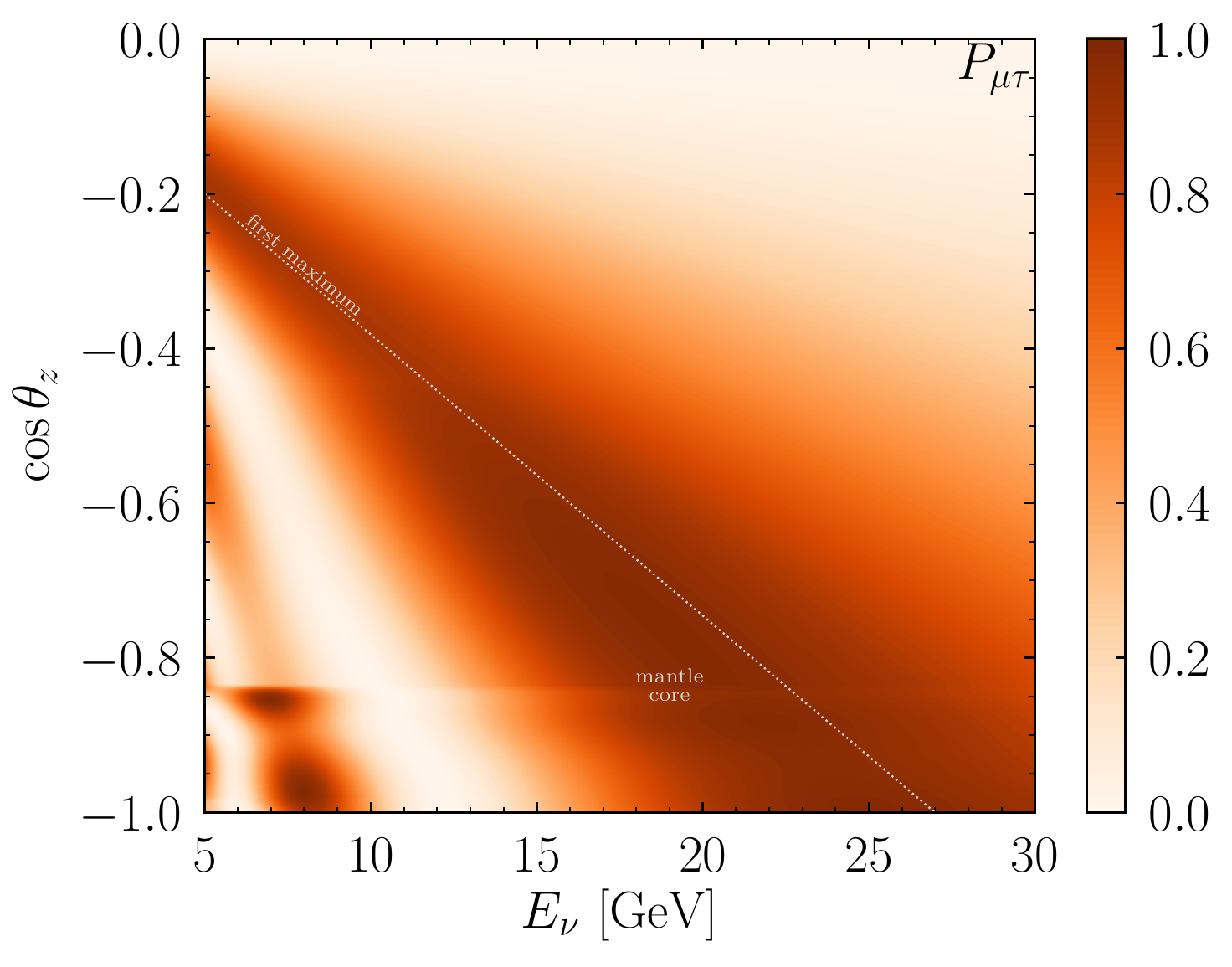}
\includegraphics[width=0.32\textwidth]{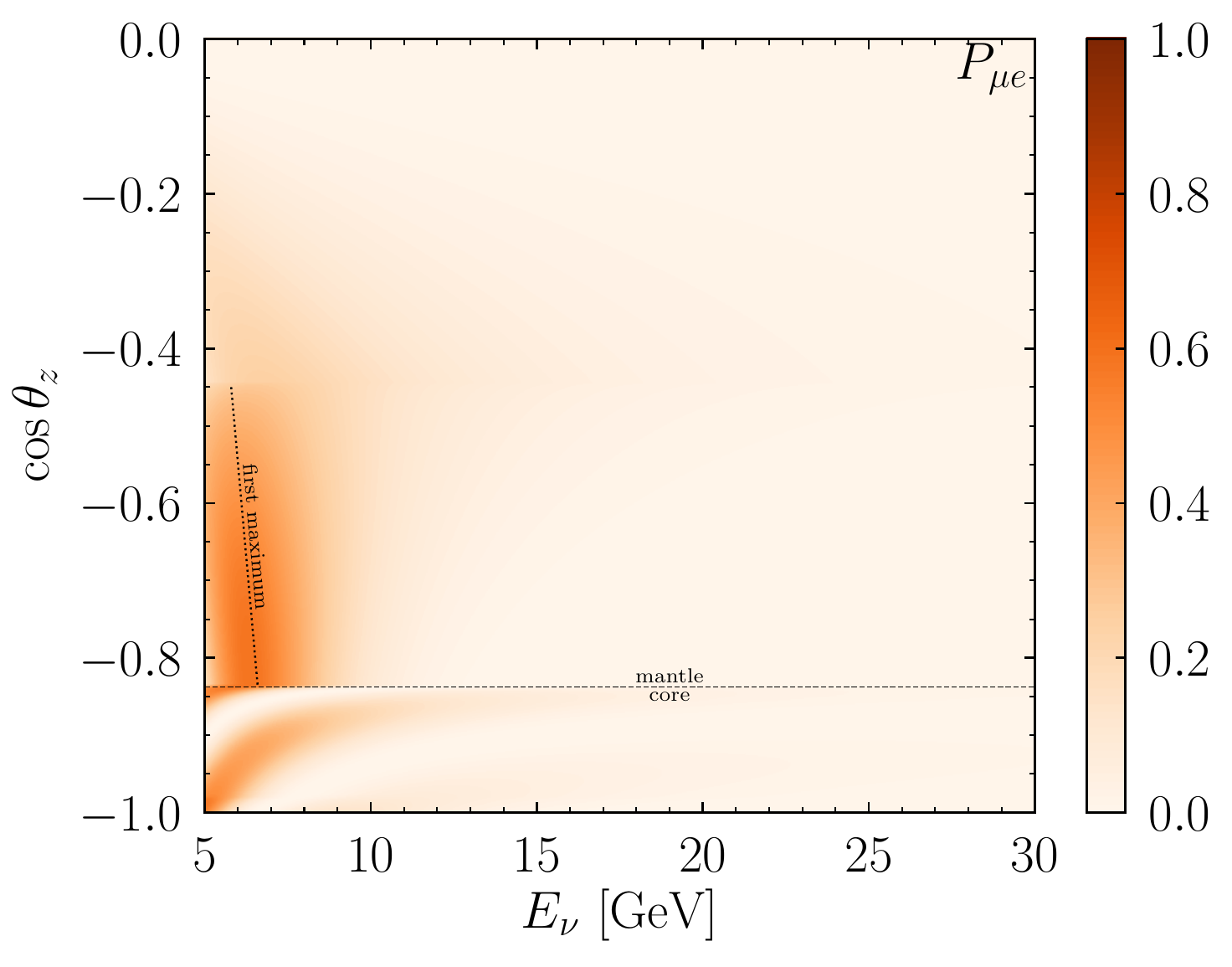}
\includegraphics[width=0.32\textwidth]{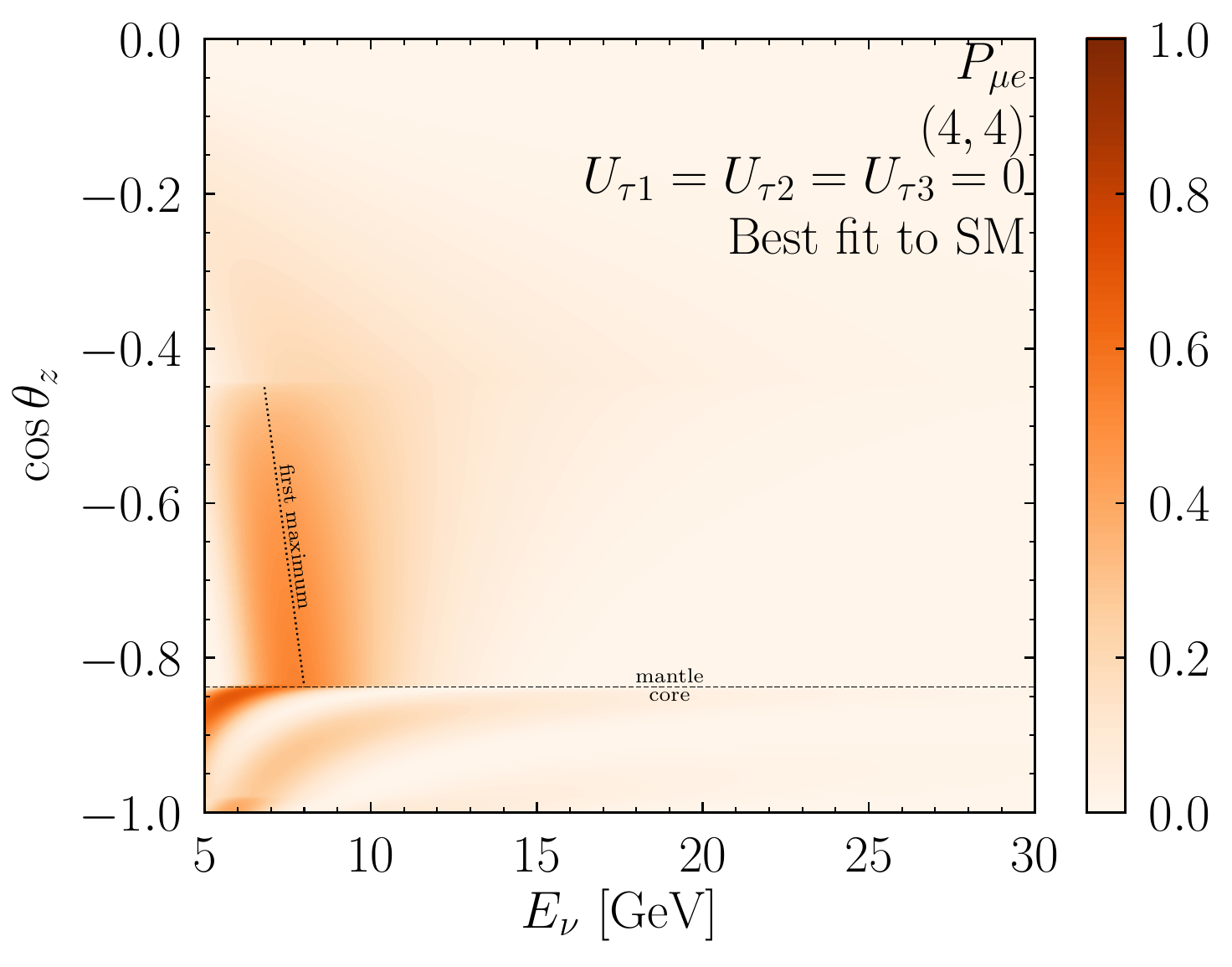}
\caption{Oscillograms of atmospheric neutrinos.
\textbf{Left}: The $\nu_\mu\to\nu_\tau$ oscillation probability for the standard oscillation parameters.
\textbf{Center}: The $\nu_\mu\to\nu_e$ oscillation probability for the standard oscillation parameters.
\textbf{Right}: The $\nu_\mu\to\nu_e$ oscillation probability in the (4,4) unitary violation case that is the best fit to the standard $\nu_\mu\to\nu_\tau$ probability as described in eq.~\ref{eq:chisq}.}
\label{fig:oscillograms}
\end{figure*}

Next, we see that in numerous different combinations of physical processes, without including either of the constraints, the sensitivity to identifying tau neutrinos is considerably enhanced.
This is somewhat counterintuitive, but can be understood in that some of the processes partially cancel.
For example, with just $R_{\tau\mu}$ (the (\n,\n,\n,\y,\n) row on the left of table \ref{tab:chisq}) the test statistic for the (5,3) case is at 57 and with just the tau neutrino reconstruction effect (the (\n,\n,\y,\n,\n) row on the left of table \ref{tab:chisq}) the test statistic is at 50, but with both (the (\n,\n,\y,\y,\n) row on the left of table \ref{tab:chisq}) the test statistic drops to 1, implying almost no sensitivity.
This is because $R_{\tau\mu}$ effectively pushes the tau neutrino appearance oscillation maximum to higher energies while $\eta_{\nu_\tau}$ reduces the reconstructed energy of the tau neutrino oscillation maximum bringing it closer to that of electron neutrinos.
In addition, further adding the $\Delta m^2_{31}$ constraint (the (\n,\n,\y,\y,\y) row on the left of table \ref{tab:chisq}) does not increase the test statistic at all since the oscillation maximum has ended up at roughly the right place so $\Delta m^2_{31}$ would be reconstructed mostly correctly.

The relationship between the NC interaction (indicated in the first columns in table \ref{tab:chisq}) and the ability to confirm there are tau neutrinos, with or without the other physical effects or the constraints, is a bit more complicated.
The reason is because not only is there a nontrivial energy dependence $\eta_{\rm NC}$ similar to the $\eta_{\nu_\tau}$ term, but there is also the fact that the $U_{\tau i}$ elements appear in the NC probability in a nontrivial way that depends on the other oscillation parameters.\\

Finally, we show the oscillograms for various interesting scenarios in fig.~\ref{fig:oscillograms}.
In the left panel we show the standard tau neutrino appearance oscillogram which is consistent with IceCube's measurements over the relevant energy and zenith angle space.
We have highlighted the primary feature, the first oscillation maximum.
We also note the presence of the core of the Earth for $\cos\theta_z<-0.84$ which significantly alters the probability at the second oscillation maximum (see e.g.~\cite{Denton:2021rgt}) but not much for the first oscillation maximum.

Next, in the center panel we show the standard oscillation case for electron neutrino appearance and note that it is significantly different.
The magnitude is quite a bit lower, although this could be compensated to reproduce the tau neutrino appearance signal by a higher flux normalization.
More importantly is the fact that the first oscillation maximum appears at much lower energies: $\sim7$ GeV compared to $\sim15$ GeV for tau neutrinos at $\cos\theta_z=-0.5$; this is driven by the matter effect which plays a considerable role in $\nu_e$ appearance and a much smaller role in $\nu_\tau$ appearance.

In the right panel we see the electron neutrino appearance channel for the best fit case without tau neutrino appearance to the standard picture in the (4,4) unitary violation scenario\footnote{The (5,3) case looks similar.}.
Notably it is quite similar to the standard case with the oscillation maximum shifted to only slightly higher energies.
This shows that, due to the effects discussed in section \ref{sec:results} it is not possible to reconstruct the standard case shown in the left oscillogram.

Throughout this paper we have assumed the normal mass ordering and have focused on neutrinos only.
The mass ordering will be well measured by DUNE and JUNO \cite{Abi:2020evt,JUNO:2021vlw} in the coming years (considering a swap in the mass ordering does not provide improvement to the fit even in cases of UV).
\emph{In principle}, one could differentiate $\nu_\tau$ from $\bar\nu_\tau$'s using the approach described here: the tau lepton production threshold and the reconstructed energy effects are very similar for each flavor, but the matter effect would induce a slight difference, especially in a UV scenario.
These effects are likely too small to be detected at present since both the antineutrino cross section and the flux are lower than for neutrinos, but perhaps next-generation experiments could attempt to separately constrain the $\nu_\tau$ and $\bar\nu_\tau$ normalizations.

We have also ignored several features such as track events from tau decays to muons as well as down-going events.
These features would only serve to provide additional information in support of the existence of tau neutrinos in an IceCube-like measurement.
Additional contributions to the flux including electron neutrinos and a very subleading tau neutrino component complicate the analysis but do not change the results.
On the other hand, some care is required as energy and angular smearing as well as topology misidentification will partially weaken the numerical results -- it is for these reasons that we caution that the exact significances may not be representative of IceCube's actual data, but the general results still apply.

\section{Conclusions}
\label{sec:conclusions}
In this paper we have investigated exactly how tau neutrinos can be identified in an atmospheric neutrino dataset at an IceCube-like experiment with no event-by-event particle identification.
We set up two different unitary-violating frameworks (4,4) and (5,3) equivalent to additional sterile neutrinos that are either kinematically accessible (the oscillation averaged 3+1 scenario), or not accessible (the $\alpha$ matrix parameterization) respectively.
These different scenarios have somewhat different consequences due to the matter effect, but the primary conclusions apply in either case.

We constructed a picture of cascade detection in atmospheric neutrinos consisting of several parts: electron neutrino charged-current events, neutral-current events, and tau neutrino charged-current events not including tau lepton decays to muons.
We also included the tau lepton production threshold and the fact that both tau neutrino events and neutral-current events deposit less energy in the detector than electron neutrino charged-current events.
We showed in fig.~\ref{fig:cascade ratio} that each of these effects has a significant change on the measured flux in both energy and zenith angle.

We then computed a test statistic comparing the scenario with no tau neutrinos to that with tau neutrinos and minimized it over the available degrees of freedom in the matrix and the flux normalization.
Consistent with expectations, without the matter effect, the tau production threshold, the tau neutrino energy reconstruction effect, or the neutral-current events, it is not possible to identify tau neutrinos, even if the electron neutrino row and $\Delta m^2_{31}$ are known.
When including all the physical effects, however, it is possible to determine that tau neutrinos can be detected in IceCube's cascade sample alone even in a unitary violating framework and without any external knowledge of the oscillation parameters.
This confirms that IceCube (as well as Super-KamiokaNDE) has \emph{definitively detected tau neutrinos in their atmospheric sample even without an assumption of unitarity}.

We encourage future analyses by IceCube, Super-KamiokaNDE, KM3NeT/ORCA, INO, and other atmospheric neutrino experiments with tau neutrino sensitivity to use the effects described here to present their results in the context of tau neutrino unitarity constraints.

\begin{acknowledgments}
We thank Carlos Arg\"uelles and Julia Gehrlein for helpful comments.
We acknowledge support from the US Department of Energy under Grant Contract DE-SC0012704.
The figures were done with \texttt{python} \cite{10.5555/1593511} and \texttt{matplotlib} \cite{Hunter:2007}.
\end{acknowledgments}

\bibliography{Tau_Unitarity_Atm}

\end{document}